\documentstyle[preprint,aps]{revtex}

\newcommand{\ket}[1]{{|#1\rangle}}
\newcommand{\bra}[1]{{\langle#1|}}
\newcommand{\calC}{{\cal C}}
\newcommand{\calP}{{\cal P}}
\newcommand{\calH}{{\cal H}}
\newcommand{\calR}{{\cal R}}
\newcommand{\calQ}{{\cal Q}}

\makeatletter
\def\@begintheorem#1#2{\par\bgroup{\it #1\ #2. }\ignorespaces}
\def\@opargbegintheorem#1#2#3{\par\bgroup{\it #1\ #2\ (#3). }%
\ignorespaces}
\def\@endtheorem{\egroup}

\def\wt{{\rm wt}}
\def\Tr{{\rm Tr}}
\def\support{{\rm supp}}
\makeatother

\newtheorem{lemma}{Lemma}
\newtheorem{theorem}{Theorem}

\begin{document}
\draft
\title{
Good quantum error-correcting codes exist.}
\author{A. R. Calderbank and Peter W. Shor}
\address{AT\&T Bell Laboratories, 600 Mountain Avenue, Murray Hill, 
NJ 07974}
\date{Submitted September 12, 1995}
\maketitle
\begin{abstract}
{
A quantum error-correcting code is defined to be a unitary mapping
(encoding) of $k$ qubits (2-state quantum systems) into a subspace
of the quantum state space of $n$ qubits such that if any $t$ of the 
qubits undergo arbitrary decoherence, not necessarily independently, 
the resulting $n$ qubits can be used to faithfully reconstruct the 
original quantum state of the $k$ encoded qubits.  Quantum 
error-correcting codes are shown to exist with asymptotic rate 
$k/n = 1-2H_2(2t/n)$ where $H_2(p)$ is the binary entropy function 
$-p \log_2 p - (1-p) \log_2 (1-p)$.  Upper bounds on this asymptotic 
rate are given.
}
\end{abstract}


\raggedbottom
\sloppy

\narrowtext

\section{Introduction}
\label{sec-intro}

With the realization that computers that use the interference and
superposition principles of quantum mechanics
might be able to solve certain problems, including prime factorization,
exponentially faster than classical computers \cite{qcalgs}, interest 
has been growing in the feasibility of these quantum computers, 
and several methods for building quantum gates and quantum computers 
have been proposed \cite{qcproposals,CiracZ}.  One of the most 
cogent arguments against the feasibility of quantum computation appears 
to be the difficulty of eliminating error caused by inaccuracy and 
decoherence \cite{qcdecoherence}.  Whereas the best experimental
implementations of quantum gates accomplished so far have less than 
90\% accuracy \cite{qcexperiments}, the accuracy required for factorization of
numbers large enough to be difficult on conventional computers
appears to be closer to one part in billions.  We hope that the
techniques investigated in this paper can eventually be extended
so as to reduce this quantity by several orders of magnitude.

In the storage and transmission of digital data, errors can be corrected 
by using error-correcting codes \cite{McWSl}.  In digital computation,
errors can be corrected by using redundancy; in fact, it has been shown 
that fairly unreliable gates could be assembled to form a 
reliable computer \cite{noisygates}. It has widely been assumed that the 
quantum no-cloning theorem \cite{noclone} makes error correction impossible 
in quantum communication and computation because redundancy cannot be 
obtained by duplicating quantum bits.  This argument was shown to be in error 
for quantum communication in Ref.~\cite{Shor2}, where a code was given that 
mapped one qubit (two-state quantum system) into nine qubits so that the 
original qubit could be recovered perfectly even after arbitrary decoherence 
of any one of these nine qubits.  This gives a quantum code on 9 qubits with 
rate $\frac{1}{9}$ that protects against one error.  
Here we show the existence of better quantum error-correcting codes, having 
higher information transmission rates and better error-correction capacity.  
Specifically, we show the existence of quantum error-correcting codes encoding
$k$ qubits into $n$ qubits that correct $t$ errors and have asymptotic rate 
$1-2H_2(2t/n)$ as $n \rightarrow \infty$.   These codes work not by 
duplicating the quantum state of the encoded $k$ qubits, but by spreading 
it out over all $n$ qubits so that if $t$ or fewer of these qubits are 
measured, no information about the quantum state of the encoded bits is 
revealed and, in fact, the quantum state can be perfectly recovered from 
the remaining $n-t$ qubits.

Suppose that we have a coherent quantum state of $k$ qubits that we 
wish to store using a physical quantum system which is subject to some 
decoherence process.  For example, during computation on the quantum computer
proposed by Cirac and Zoller \cite{CiracZ},
we would need to store quantum information in entangled 
electronic states of ions held in an ion trap.
The decoherence time of the quantum state of $k$ 
entangled qubits is in general $1/k$ of the decoherence time of 
one qubit (this makes the optimistic assumption that coherence between 
different qubits is as stable as coherence of a single qubit).
Thus, one might expect that the best way to store the state of $k$
entangled qubits is to store them in $k$ physical qubits.  Our results
show that if we use quantum error-correcting codes, it is possible
to store the $k$ qubits in $n>k$ qubits so that the decoherence time 
for the encoded quantum state is a small constant fraction of the 
decoherence time of one qubit.
These results thus show that some measurable non-local properties of 
entangled systems are much more stable under decoherence than is the 
entire entangled system.

Physical quantum channels will be unlikely to leave $n-t$ qubits 
perfectly untouched and 
subject the remaining $t$ qubits to decoherence.  To analyze the behavior 
of our error-correcting code for physical quantum channels,
we must make some assumptions about the decoherence process.  In 
Section~\ref{sec-q-channels}, we will show that our error correction 
method performs well if the decoherence of different qubits 
occurs independently, i.e., if each of the qubits is coupled to 
a separate environment.  Our error-correction method will actually work 
for more general channels, as it can tolerate coupled decoherence 
behavior among small groups of qubits.  

The lower bound of $1-2H_2(2t/n)$ shown in our paper should be compared 
with theoretical upper bounds of 
\[
\min\left[1-H_2(2t/3n),\linebreak[3] 
H_2\left(\textstyle\frac{1}{2} + \sqrt{(1-t/n)t/n}\,\right)\right]
\] 
for $t/n < \frac{1}{2}$, and 
$0$ for $t/n \geq \frac{1}{2}$.  These are obtained from bounds on the 
quantum information capacity of a quantum channel, which we derive in
Section \ref{sec-q-channels} from 
results of Refs. \cite{levkh,BeBrScSmWo}.  These bounds are plotted
in Fig.~\ref{qcplots} in Section \ref{sec-q-channels}.

\section{Definitions}
\label{sec-defs}

Our constructions of quantum error-correcting codes rely heavily on
the properties of classical error-correcting codes.  We will thus 
first briefly review certain definitions and properties related to 
binary linear error-correcting codes.  We only consider vectors and
codes over ${\sf F}_2$, the field of two elements, so we have
$1+1=0$.  A binary vector $v \in {\sf F}_2$ with $d$ 1's is said to 
have {\it Hamming weight} $d$, denoted by $\wt(v)=d$.  
The {\it Hamming distance} $d_H(v,w)$
between two binary vectors $v$ and $w$ is $\wt(v+w)$.  
The {\it support} of a vector $v$, denoted by $\support(v)$,
is the set of coordinates of $v$ where the corresponding entry is 
not 0, that is, $\support(v)=\{i:v_i \ne 0\}$.  Suppose that $S$ is 
a set of coordinates.  Then $v|_{S}$ denotes the projection of $v$ 
onto $S$, i.e., the vector that agrees with $v$ on the coordinates 
in $S$ and is $0$ on the remaining coordinates.  For a binary vector 
$E$ we use $v|_E$ to mean $v|_{\support(E)}$.  We also use 
$e \preceq E$ to mean that $\support(e) \subseteq \support(E)$.

A {\it code} $\calC$ of length $n$
is a set of binary vectors of length $n$, called {\it codewords}.
In a {\it linear code} the codewords are those vectors in
a subspace of ${\sf F}_2^n$ (the $n$-dimensional vector 
space over the field ${\sf F}_2$ on two elements).  The {\it minimum 
distance} $d=d(\calC)$ of a binary code $\calC$ is the minimum 
distance between two distinct codewords.  If $\calC$ is linear then 
this minimum distance is just the minimum Hamming weight of a 
nonzero codeword.

A linear code with length $n$, 
dimension $k$, and minimum distance $d$ is called an $[n,k,d]$ code.
For a code $\calC$ with minimum distance $d$,
any binary vector in ${\sf F}_2^n$ is within Hamming distance  
$t=\lfloor \frac{d-1}{2} \rfloor$ of at most one codeword; thus, a
code with minimum distance $d$ can correct $t$ errors made in the bits
of a codeword; such a code is thus said to be a $t$ error-correcting 
code.  The {\it rate} $R$ of a 
linear code of length $n$ is $\dim(\calC)/n$; this is the ratio of 
the information content of a codeword to the information content of
an arbitrary string of length $n$.  The {\it dual code} 
$\calC^\perp$ of a code $\calC$ is the set of vectors of 
perpendicular to all codewords, that is, 
$\calC^\perp = \{ v \in {\sf F}_2^n : v\cdot c = 0 \ \forall c 
\in \calC \}$.  From linear algebra, 
$\dim(\calC)+\dim(\calC^\perp) = n$.

In this paper, we will use the $[7,4,3]$ Hamming code as an example to
illustrate our construction of quantum error-correcting codes.  
This code contains the following 16 binary vectors of length 7:
\begin{equation}
\begin{array}{cccc}
0000000, & 0001011, & 0010110, & 0011101, \nonumber \\
0100111, & 0101100, & 0110001, & 0111010, \label{Hamm} \\
1000101, & 1001110, & 1010011, & 1011000,  \nonumber \\
1100010, & 1101001, & 1110100, & 1111111.   \nonumber
\end{array}
\end{equation}
The minimum distance is the minimum Hamming weight of a non-zero 
codeword, which is 3, so this is a one-error correcting code.
It is easily verified that the dual code consists of all vectors in the 
Hamming code with an even weight.

The {\it quantum Hilbert space} $\calH_2^n$ over $n$ qubits is the 
complex space generated by basis vectors $\ket{b_0}$, $\ket{b_1}$, 
$\ldots$, $\ket{b_{2^n-1}}$ where $b_i$ is the representation of the 
number $i$ in binary.  This Hilbert space has a natural 
representation as a tensor product of $n$ copies of $\calH_2$, 
with the $i$th copy corresponding to the $i$th bit of the basis 
vectors.  We refer to each of these copies of $\calH_2$ as 
a {\it qubit}.

We define a {\it quantum error-correcting code} $\calQ$ with rate $k/n$ 
to be a unitary mapping of $\calH_2^{k}$ into $\calH_2^n$.
Strictly speaking, this is actually a unitary mapping of $\calH_2^k$ into a 
$2^k$-dimensional subspace of $\calH_2^n$; it can alternatively be viewed
as a unitary mapping of $\calH_2^{k}\otimes \calH_2^{n-k}$ into
$\calH_2^n$, where the quantum state in $\calH_2^{n-k}$ is taken to
be that where all the qubits have quantum state $\ket{0}$.  
In our model of error analyzed in Section \ref{sec-decoding}, we will
assume that the decoherence process affects only $t$ bits; that is, 
the decoherence is modeled by first applying an arbitrary unitary 
transformation $D$ to the space consisting of the tensor product 
$\calH_2^t \otimes \calH_{\rm env}$
of any $t$ of the qubits and some arbitrary Hilbert space 
$\calH_{\rm env}$ designating the environment, and then tracing over the
environment $\calH_{\rm env}$ to obtain the output of the channel, which 
will thus in general be an ensemble of states in $\calH_2^{k}$.
We say that a quantum code can correct $t$ errors if
the original state $\ket{x} \in \calH_2^{k}$
can be recovered from the decohered encoded state $D\calQ\ket{x}$
by applying a unitary transformation $\calR$ (independent of $D$) to 
$\calH_2^n \otimes \calH_{\rm anc}$, where $\calH_{\rm anc}$ is a 
Hilbert space representing the state of an ancilla (i.e., a 
supplementary quantum system).
It turns out that if our quantum code will correct an 
{\it arbitrary} decoherence of $t$ or fewer qubits, it will 
also be able transmit information with high fidelity for a large class
of channels with physically plausible decoherence processes;
this is discussed in Section \ref{sec-q-channels}.

Since the error correction must work for any encoded state 
$\calQ\ket{x}$, the property of being a quantum error-correcting 
code depends only on the subspace $\calQ\calH_2^k$ of $\calH_2^n$,
and not on the actual mapping $\calQ$.  However, for
ease of explanation, we will nonetheless define an orthogonal
basis of this subspace of $\calH_2^n$, which can be used to obtain
an explicit mapping $\calQ$, and call the elements of this basis 
codewords.  

\section{Quantum Codes}
\label{sec-q-codes}

We will now define our quantum code.
Suppose that we have a linear code $\calC_1 \subset {\sf F}_2^n$.  
We let $\calH_{\calC_1}$ be the subspace of $\calH_2^n$ generated by
vectors $\ket{c}$ with $c \in \calC_1$.
Let $M$ be a generator matrix for $\calC_1$; this means that 
$\calC_1$ is the row space of $M$, so that $vM$ ranges over all
the codewords in $\calC_1$ as $v$ ranges over all vectors in
${\sf F}_2^{\dim(\calC_1)}$.  
For $w \in {\sf F}_2^n$, we define a quantum state $\ket{c_w}$ by
\begin{equation}
\ket{c_w} = 2^{-\dim(\calC_1)/2}
\sum_{v \in {\sf F}_2^{\dim(\calC_1)}} (-1)^{vMw}\ket{vM}.
\label{defcw}
\end{equation}
Note that if $w_1 + w_2 \in \calC_1^\perp$, then $\ket{c_{w_1}} = 
\ket{c_{w_2}}$, since $vMw_1 = vMw_2$ for all 
$v \in {\sf F}_2^{\dim(\calC_1)}$.
Further note that $\langle c_{w_1} | c_{w_2} \rangle = 0$ if
$w_1 + w_2 \not\in \calC_1^\perp$. This follows since 
$\sum_{v} (-1)^{vMw} = 0$ unless $vMw=0$ for all $v \in F_2^{\dim(\calC_1)}$
Thus, for 
$w \in {\sf F}_2^n / \calC_1^\perp$ 
the vectors $\ket{c_w}$ form a basis for the
space $\calH_{\calC_1}$.  (Here $F_2^n/{\calC_1^\perp}$ stands for the
cosets of $\calC_1^\perp$ in ${\sf F}_2^n$, which are the sets
$\calC_1^\perp + w$ where $w \in {\sf F}_2^n$; there are 
$2^{\dim(\calC_1)}$ of these cosets and they form the 
natural index set for the quantum states $\ket{c_w}$.)

Suppose now that we have another linear code $\calC_2$ with
$\{0\} \subset \calC_2 \subset \calC_1 \subset {\sf F}_2^n$.  
Our quantum code will be constructed using codes $\calC_1$ and $\calC_2$.
We define the codewords of our quantum code 
$\calQ_{\calC_1,\calC_2}$ as the set of $\ket{c_{w}}$ for 
all $w \in \calC_2^\perp$,
Recall that two codewords
$\ket{c_w}$ and $\ket{c_{w'}}$ are equal if $w+w' \in \calC^\perp$.
The natural index set for the codewords is thus over
$\calC_2^\perp /\calC_1^\perp$,
the cosets of  $\calC_1^\perp$ in $\calC_2^\perp$.
This code thus contains $2^{\dim(\calC_1)-\dim(\calC_2)}$ orthogonal
vectors. Since its length is $n$ qubits, it has  
rate $(\dim(\calC_1)-\dim(\calC_2))/n$.  To construct a quantum 
error-correcting code from the Hamming code given in
Eq.~(\ref{Hamm}), we will take $\calC_1$ to be this code and
$\calC_2$ to be $\calC_1^\perp$.  Thus, $\dim(\calC_1)=4$ and 
$\dim(\calC_2)=3$, so our quantum error-correcting code will map $4-3=1$ 
qubit into 7 qubits.  There are thus two codewords.  The first is
\begin{eqnarray}
\ket{c_0} &=& 
\frac{1}{4}\, \Big(\ \ \ket{0000000} + \ket{0011101} +
\ket{0100111} + \ket{0111010} \nonumber \\
& &\ \ \ \ +\ket{1001110}+ \ket{1010011} 
+ \ket{1101001} + \ket{1110100} \\
& &\ \ \ \ + \ket{0001011} + \ket{0010110}  
+ \ket{0101100}  +  \ket{0110001} \nonumber \\
& &\ \ \ \  + \ket{1000101} +\ket{1011000} + \ket{1100010} 
 + \ket{1111111} \ \Big)\nonumber \,,
\end{eqnarray}
and the second is
\begin{eqnarray}
\ket{c_1} &=& 
\frac{1}{4}\, \Big(\ \ \ket{0000000} + \ket{0011101} +
\ket{0100111} + \ket{0111010} \nonumber \\
& &\ \ \ \  +\ket{1001110}+ \ket{1010011} 
+ \ket{1101001} + \ket{1110100} \\
& &\ \ \ \  - \ket{0001011} - \ket{0010110}  
- \ket{0101100}  -  \ket{0110001} \nonumber \\
& &\ \ \ \ - \ket{1000101}-\ket{1011000} - \ket{1100010} 
 - \ket{1111111} \ \Big)\nonumber \,.
\end{eqnarray}
Note that in $\ket{c_1}$ all the codewords of the Hamming code with an odd weight 
have a negative amplitude, and all the codewords with even weight have 
positive amplitude.  This is the effect of the $(-1)^{vMw}$ term in 
Eq.~(\ref{defcw}).

We will show that if $\calC_1$ and $\calC_2^\perp$ have minimum distance $d$, 
then the quantum code $\calQ_{\calC_1,\calC_2}$ can correct 
$t = \lfloor\frac{d-1}{2}\rfloor$ errors.  (For our example code,
$\calC_1 = \calC_2^\perp$ has minimum distance 3, 
so our quantum code will correct one error.)
In the remainder of this section,
we will give some intuition as to why this should be true; while in the 
next section, we will work out this calculation in detail.

To show why our codes are error-correcting, we must first give 
another representation of our codewords.  If we perform the following
change of basis,
\begin{eqnarray}
\ket{0} & \rightarrow & {\textstyle\frac{1}{\sqrt{2}}} (\ket{0}+ \ket{1}) 
\label{basis-change}\\
\ket{1} & \rightarrow & {\textstyle\frac{1}{\sqrt{2}}} (\ket{0}- \ket{1}), 
\nonumber
\end{eqnarray}
to each of the bits of our codeword $\ket{c_w}$ we obtain the state
\begin{equation}
\ket{s_w} = 2^{(\dim(\calC_1)-n)/2} \sum_{u \in \calC_1^\perp} \ket{u+w}.
\label{defsw}
\end{equation}
We can see this since if $\ket{x}$ is any basis state in the rotated
basis given by Eq.~(\ref{basis-change}), then 
\begin{equation}
\langle x | c_v \rangle = 2^{-(n+\dim{\calC_1})/2}
\sum_{v \in {\sf F}_2^{\dim(\calC_1)}}
(-1)^{vM(w+x)},
\end{equation}
and this sum is 0 unless $w+x \in \calC_1^\perp$.
Letting $u = w+x$, we get Equation (\ref{defsw}).
For our example quantum code, 
\begin{eqnarray}
\ket{s_0} &=&
\frac{1}{2\sqrt{2}}\, \Big(\ \ \ket{0000000} + \ket{0011101} +
\ket{0100111} + \ket{0111010}  \\
& &\ \ \ \ \ \ \ \  +\ket{1001110}+ \ket{1010011}
+ \ket{1101001} + \ket{1110100} \ \Big) \nonumber
\end{eqnarray}
and
\begin{eqnarray}
\ket{s_1} &=&
\frac{1}{2\sqrt{2}}\, \Big(\ \ \ket{0001011} + \ket{0010110}
+ \ket{0101100}  +  \ket{0110001} \\
& &\ \ \ \ \ \ \ \ + \ket{1000101}+\ket{1011000} + \ket{1100010}
 + \ket{1111111} \ \Big)\nonumber \,.
\end{eqnarray}

We can now see how these codes are able to correct errors.  In the $\ket{c_w}$
representation, all the codewords are superpositions of basis
vectors $\ket{v}$ with $v \in \calC_1$.  Thus, any $t$ bit errors 
(those errors taking $\ket{0} \rightarrow \ket{1}$ and 
$\ket{1} \rightarrow \ket{0}$) 
can be corrected by performing a classical error-correction 
process for the code
$\calC_1$ in the original basis.  In the $\ket{s_w}$ representation, 
all the codewords are superpositions
of basis vectors $\ket{v}$ with $v \in \calC_2^\perp$.  Thus, any $t$ bit
errors in the rotated basis can be corrected by performing a classical 
error-correction process for the code $\calC_2^\perp$
in the rotated basis.  However, phase errors in
the original basis (errors taking $\ket{0} \rightarrow  \ket{0}$ and
$\ket{1} \rightarrow - \ket{1}$) are bit errors in the rotated basis
and vice versa.  Thus, our quantum code can correct $t$ bit errors and
$t$ phase errors in the original basis.

The correction process we use for our quantum error-correcting codes
is indeed to first correct bit errors in the $\ket{c_v}$ basis 
classically and then to correct bit errors in the $\ket{s_v}$ basis
classically.  It remains to be shown that the correction process 
for the bit errors does not interfere with the correction process 
for the phase errors, and that arbitrary non-unitary errors on $t$ or
fewer quantum bits of our code will also be corrected by this procedure.  
This is done through calculations which are performed in 
Section \ref{sec-decoding} of our paper.

As in Ref.~\cite{Shor2}, we correct the error by measuring the
decoherence without disturbing the encoded information.  Intuitively,
what we do is to measure the decoherence without observing
the encoded state; this then lets us correct the decoherence while
leaving the encoded state unchanged.  In our decoding procedure,
we thus learn which qubits had bit errors and which had phase errors,
which tells us something about the decoherence 
process but which gives no information about our encoded state.
Linear codes are very well suited for this application: each codeword
has the same relation to all the other words in the code, and this
property is what enables us to measure the error without learning 
which codeword it is that is in error.

Since this paper was submitted, we learned that related work has been
done by Steane \cite{Steane}.  
Steane generates his quantum code using codewords
\begin{equation}
\ket{s'_{w}} = 2^{-\dim(\calC_2)/2}  \sum_{v \in \calC_2} \ket{v+w},
\end{equation}
where $w$ is chosen from $\calC_1 / \calC_2$.  This is the same as our
$\ket{s_w}$ basis if the codes $\calC_1$ and $\calC_2^\perp$ are
interchanged.  It should also be noted that these codewords $\ket{s'_{w}}$
generate exactly the same subspace of $\calH_2^n$ as the codewords 
$\ket{c_w}$ given by Eq.~\ref{defcw}, and thus effectively give a different
basis for the same quantum code.

\section{Decoding Quantum Codes}
\label{sec-decoding}

In this section we will show that errors in any $t$ qubits of
our quantum codes can be 
corrected by first correcting bit errors in the $\ket{c}$ basis, and
then correcting bit errors in the $\ket{s}$ basis.  
For this section and the remainder of this paper, we will assume for 
simplicity that $\dim(\calC_1) = n-k$ and $\dim(\calC_2) = k$; thus,
the rate of our codes will be $1-2k/n$.  However,
all of our results are easily extendable to quantum codes derived from
classical codes $\calC_2 \subset \calC_1 \subset {\sf F}_2^n$ of 
any dimension.

In order to prove that errors in quantum codes can be corrected, we 
first need a lemma about purely classical codes.
\begin{lemma}
Suppose that $\calC$ is a binary linear code of length $n$.  
Let $e$, $E$ $\in$ ${\sf F}_2^n$, with $e \preceq E$ and
$\wt(E) < d(\calC^\perp)$.   Then 
there exists a vector $v_e \in \calC$ such that 
$v_e|_{\support(E)} = e$.  \label{inverselemma}
\end{lemma}
\begin{proof}
The projection of $\calC$ onto $E$ has to have full rank, because 
otherwise $\calC^\perp$ would contain a vector $w$ with 
$\wt(w) \leq \wt(E) < d(\calC^\perp)$.
\end{proof}

We now need the following lemma about the
states $\ket{c_w}$.

\begin{lemma}
Suppose that $\calC_1$ has minimum distance $d$.  Let $e$, $E$ $\in$ 
${\sf F}_2^n$ with $e \preceq E$.  
Let $P$ be the projection onto the subspace
of $\calH_2^n$ generated by all $\ket{v}$ where $v$ is in the set
$\{v \in {\sf F}_2^n: v|_E = e\}$, 
that is, with $v$ equal to $e$ on $\support(E)$.  Then 
\begin{mathletters}
\begin{eqnarray}
\left\langle c_{w_1} | P | c_{w_2} \right\rangle & = &
2^{-(n-k)} \sum_{v: vM|_E = e} (-1)^{vM(w_1+ w_2)} 
\label{lemmasum}\\
& = & \left\{ 
\begin{array}{ll}
{(-1)^{e\cdot (c+w_1+ w_2)}}/{2^{\wt(E)}} &
\hbox{ if }\exists c \in \calC_1^\perp \hbox{ such that } 
c + w_1 + w_2 \preceq E, \\
0  &
\hbox{ otherwise.}
\end{array}
\right.
\label{lemmaeq}
\end{eqnarray}
\end{mathletters}%
\end{lemma}

\begin{proof} From the definition of $\ket{c_w}$ in Eq.~(\ref{defcw}), it
is straightforward to show Eq.~(\ref{lemmasum}).
We must now show that this is equal to Eq.~(\ref{lemmaeq}).
Since $\wt(e) < d(\calC_1^\perp)$, by Lemma 1
there is a vector $v_e$ such that $v_eM|_E = e$.
We can obtain the linear space $\{v\in {\sf F}_2^{n-k} : v|_E=e\}$ 
by taking every vector in the set 
$\{v\in {\sf F}_2^{n-k} : v|_E =0\}$ 
and adding the vector $v_e$.  Using this substitution in 
Eq.\ (\ref{lemmasum}) gives
\begin{mathletters}
\begin{eqnarray}
\left\langle c_{w_1} | P | c_{w_2} \right\rangle &=&
2^{-(n-k)}\sum_{v : vM|_E=0}
(-1)^{(v + v_e)M(w_1 + w_2)} \\
& = & 
2^{-(n-k)}(-1)^{v_eM(w_1 + w_2)}
\sum_{v : vM|_E=0}
(-1)^{vM(w_1  +  w_2)} .
\label{Esum}
\end{eqnarray}
\end{mathletters}%

Now, because the set $\{vM: vM|_E=0\}$ is an $n-k-\wt(E)$ dimensional
subspace of ${\sf F}_2^k$, the
sum (\ref{Esum}) is 0 unless $vM(w_1 + w_2)=0$ for all $vM$ in this
subspace.  It is clear that if there is a $c \in \calC_1^\perp$ such 
that $w_1+ w_2 + c \preceq E$, then $vM(w_1 + w_2) = 0$ if
$vM|_E=0$, and $v_eM(w_1 + w_2) = e \cdot (c+w_1+w_2)$.  This shows
the first part of Eq.~(\ref{lemmaeq}).

We now prove the other direction.  Suppose
that $vM(w_1 + w_2) = 0$ for all $v$ with $vM|_E=0$.  
Let $e_j$ be the vector that is $1$ on the $j$th coordinate
of $E$ and $0$ on the other coordinates.  
We know from Lemma \ref{inverselemma} that there is
a vector $v_j \in {\sf F}_2^{n-k}$ such that $v_j M |_E = e_j$.  
Let $\sigma_j = v_jM(w_1 + w_2)$.  We consider the vector 
$c' = w_1 + w_2 + \sum_{j=1}^{\wt(E)} \sigma_j e_j$;
we will show that this vector satisfies the conditions for the $c$ 
in Eq.~(\ref{lemmaeq}).  
Clearly, $w_1 + w_2 + c' \preceq E$. We need also to show that
$c' \in \calC_1^\perp$.
Consider any vector 
$v \in {\sf F}_2^{n-k}$.  We can decompose it into 
$v = v_0 + \sum_{i=1}^{\wt(E)} \alpha_i v_i$ where $v_0M|_E = 0$, and
$\alpha_i$ is 0 or 1.  Note that $v_i M e_j = \delta(i,j)$ where 
$\delta$ is the Kronecker delta function.  Now, 
\begin{mathletters}
\begin{eqnarray}
v M c' &=& (v_0 + \sum_{i=1}^{\wt(E)} \alpha_i v_i) M
( w_1 + w_2 + \sum_{j=1}^{\wt(E)} \sigma_j e_j) 
\label{vMceq} \\
&=& (\sum_{i=1}^{\wt(E)} \alpha_i v_i) M
( w_1 + w_2 + \sum_{j=1}^{\wt(E)} \sigma_j e_j) \\
&=& \sum_{i=1}^{\wt(E)} \alpha_i v_i M (w_1+ w_2) + 
\sum_{i=1}^{\wt(E)} \alpha_i \sigma_i
\label{twoterms} \\
&=& 0, \nonumber
\end{eqnarray}
\end{mathletters}%
proving the second part of Eq.~(\ref{lemmaeq}).
The terms containing $v_0$ vanish in Eq.~(\ref{vMceq}) because 
 $v_0M(w_1+w_2) = 0$ since $v_0M|_E = 0$, and
$v_0Me_i = 0$ since $e_i \prec E$.  The two terms in 
Eq.~(\ref{twoterms}) cancel because of the definition of $\sigma_i$.
\end{proof}

We are now ready to prove the following theorem.

\begin{theorem}
If $\calC_1$ and $\calC_2^\perp$ are both linear $[n,n-k,d]$ codes 
with $\{0\} \subset \calC_2 \subset \calC_1 \subset {\sf F}_2^n$, 
then the quantum code $\calQ_{\calC_1,\calC_2}$
is a $t$-error correcting code, where 
$t = \lfloor\frac{d-1}{2}\rfloor$.
\end{theorem}

\begin{proof}
We show how to correct any $t$ errors.  Let us start with
a codeword $\ket{c_w}$ for $w \in \calC_2^\perp$.  Now, let 
$E$ be the binary vector such that $\support(E)$ is the set 
of qubits that have decohered.  By our hypothesis that at most $t$ 
qubits decohere, we can take $\wt(E) = t$.
We denote states of the environment by $\ket{a_i}$.
Since the decoherence only operates on those qubits in 
$\support(E)$, the most general decoherence $D$ is a unitary 
process operating on a binary vector $u$ and the initial state 
of the environment $\ket{a_0}$ as follows:
\begin{equation}
D \ket{u,a_0} = \sum_{e \preceq E}
\ket{u + e} \ket{a_{u|_E,e}},
\end{equation}
where the states of the environment $|a_i\rangle$ are not necessarily
normalized.
Now, we let this decoherence act on $\ket{c_w}\ket{a_0}$.  We get
\begin{equation}
D\ket{c_w,a_0} = 2^{-(n-k)/2}\sum_{v\in{\sf F}_2^{n-k}}
(-1)^{vMw} \sum_{e \preceq E} \ket{vM+ e} \ket{a_{vM|_E,e}}.
\end{equation}
Now, we know $vM \in \calC_1$, which is a code with 
minimum distance $d > 2\wt(e)$.  Thus, we can restore $vM+ e$
to a unique codeword $vM \in \calC_1$.  Intuitively, this corrects 
bits that have flipped from $0$ to $1$ or {\it vice versa}.  
We can do this using a 
unitary operator $\calR_f$ provided we make the operation reversible;
to do this we record the error $e$
in a set of ancilla qubits $A$.  After this process, the quantum
state of our system is
\begin{equation}
\calR_f D \ket{c_w} = 
2^{-(n-k)/2}
\sum_v (-1)^{vMw} \sum_{e \preceq E} 
\ket{vM} \ket{a_{vM|_E,e}}\ket{A_e}.
\label{afterflip}
\end{equation}
Note that since $vM \in \calC_1$, we have now corrected our state to
some state in the Hilbert space $\calH_{\calC_1}$.  Recall that
the vectors $\ket{c_u}$ with 
$u \in {\sf F}_2^n$ generated $\calH_{\calC_1}$.
What we do now is to consider the Hilbert space $\calH_{\calC_1}$ 
in terms of the basis elements $\ket{c_u}$ for 
$u \in {\sf F}_2^n/\calC_1^\perp$ 
instead of the basis elements $\ket{vM}$.  We do this by 
substituting the identity
\begin{equation}
\ket{vM} = 
2^{-(n-k)/2}
\sum_{u \in {\sf F}_2^n/ \calC_1^\perp} (-1)^{vMu} \ket{c_u}
\label{second-change}
\end{equation}
in Eq.~(\ref{afterflip}).  This gives the same type of 
effect as the change of basis in Eq.~(\ref{basis-change}) 
in that it produces a representation in which it is easier 
to deal with phase errors.
The substitution (\ref{second-change}) gives the equation
\widetext
\begin{equation}
\calR_f D \ket{c_w} = 
2^{-(n-k)}
\sum_v (-1)^{vMw} \sum_u (-1)^{vMu} \ket{c_u} \sum_{e\preceq E}
\ket{a_{vM|_E,e}} \ket{A_e},
\end{equation}
which can be rewritten as
\begin{equation}
\calR_f D \ket{c_w} = 
2^{-(n-k)}
\sum_{e \preceq E} \ket{A_e} \sum_{e' \preceq E} \ket{a_{e',e}} 
\sum_{u} \ket{c_u} \sum_{v:vM|_E=e'} (-1)^{vMw} (-1)^{vMu} 
\label{beforephases}
\end{equation}
\narrowtext
Now, by Lemma~2, the inner sum is 0 unless 
there exists $c \in {\calC_1^\perp}$ for which
$c+w+ u \preceq E$.  
This means that $\ket{c_w}$ can only decohere to $\ket{c_u}$ if 
there is a $c \in \calC_1^\perp$ such that $\wt(u + w + c) \leq t$.  
We now show this means that for each $\ket{c_u}$ there is a 
unique $\ket{c_w}$ with $w \in \calC_2^\perp / \calC_1^\perp$ 
which it could have arisen from.  Suppose that we have two 
such $w$'s, $w_1$ and $w_2$ with $w_1 + u + c_1 = e_1$ and 
$w_2 + u + c_2 = e_2$.
Then,
\begin{equation}
e_1 + e_2 = w_1 + w_2 + c_1 + c_2 \in \calC_2^\perp.
\end{equation}
However,
\begin{equation}
\wt(e_1+e_2) \leq \wt(e_1) + \wt(e_2) \leq 2t.
\end{equation}   
But $\calC_2^\perp$ has minimum distance $d > 2t$; thus $e_1 = e_2$, so
$w_1+w_2 \in C_1^\perp$ and $\ket{c_{w_1}} = \ket{c_{w_2}}$.  

This means that we can unitarily express
the state in Eq.~(\ref{beforephases}) in terms of $\ket{c_u}$, where 
$u \in {\sf F}_2^n
/ C_1^\perp$, and then correct the state $\ket{c_u}$ to $\ket{c_w}$,
since there is at most one $w$ with $d_H(w,u) < t$.
As before, to unitarily correct $\ket{c_u}$ to $\ket{c_w}$
we need to use a second ancilla $A'$ to 
record which bits we needed to flip to get from $u$ to $w$.  
These flipped bits correspond to phase errors in the original 
basis.  Denoting this correction operator by $\calR_p$, we get
\begin{eqnarray}
\calR_p \calR_f D \ket{c_w} & = &
2^{-(n-k)}
\sum_{e \preceq E} \ket{A_e} \sum_{e' \preceq E} \ket{a_{e',e}} 
\sum_{v:vM|_E=e'} \sum_{e'' \preceq E} 
(-1)^{vMw} (-1)^{vM(w + e'')} \ket{c_w} \ket{A'_{e''}}  \nonumber \\
& = &
2^{-(n-k)}
\ket{c_w} \sum_{e \preceq E} \ket{A_e} \sum_{e' \preceq E} 
\ket{a_{e',e}} \sum_{e'' \preceq E} \ket{A_{e''}}\sum_{v:vM|_E=e'} 
(-1)^{vMe''} \\
&=&
2^{-\wt(E)}
\ket{c_w} \sum_{e \preceq E} \ket{A_e} 
\sum_{e'' \preceq E} \ket{A'_{e''}} 
\sum_{e' \preceq E} (-1)^{e'\cdot e''}\ket{a_{e',e}}, \nonumber
\end{eqnarray}
which is just $\ket{c_w}$ tensored with a state of the ancillae and 
the environment that {\em does not depend}
on $w$.  We have thus unitarily restored the original state and 
corrected $t$ decohered bits.
\end{proof}

\section{Weakly Self-Dual Codes}
\label{sec-cl-codes}

To show that a family of codes contains codes that meet the 
Gilbert-Varshamov bound we can often employ a very simple greedy 
argument; this argument appears in Ref.~\cite{McWSl}, pp.\ 557--558 
(proof of Thm.~31 of Chap.~17).
\begin{lemma}
Let $\phi_i$ be a set of $[n_i, k_i]$ codes such that
\begin{enumerate} 
\item $k_i/n_i >R$
\item each nonzero vector of length $n_i$ belongs to the same 
number of codes in $\phi_i$.
\end{enumerate} 
Then there are codes in the family that asymptotically meet the 
Gilbert-Varshamov bound:
\begin{equation}
R \geq 1-H_2({\textstyle\frac{d}{n}}) \hbox{ \ \ \ as } 
n \rightarrow \infty
\end{equation} 
\end{lemma}
\begin{proof}
Let $W_i$ be the number of codes in $\phi_i$ that contain a 
particular vector $v$.  By hypothesis,
\begin{equation}
(2^{n_i} -1) W_i = (2^{k_i}-1) |\phi_i|.
\end{equation}
The number of vectors with weight less than $d$ is 
\begin{equation}
\sum_{j=0}^{d-1}{n_i \choose j}.
\end{equation}
If 
\begin{equation}
W_i \sum_{j=0}^{d-1} {n_i\choose j} < 
W_i(2^{n_i} -1)/ (2^{k_i}-1) = |\phi_i|
\end{equation}
then there is a code in $\phi_i$ with minimum distance $\geq d$.
\end{proof}

This proof is not constructive in that it does not produce codes
satisfying this bound, but merely shows that they exist.  In fact,
explicit constructions for classical codes that attain the 
Gilbert--Varshamov bound asymptotically are not known.

Consider towers of codes as shown below: 
\begin{equation}
\{0\} \subseteq \langle\!\langle 1^n \rangle\!\rangle\subseteq \calC 
\subseteq \calC^\perp
\subseteq F_2^n
\end{equation}
where $\dim{\calC} = k$ and $\dim{\calC^\perp} = n-k$.  Here 
$\langle\!\langle 1^n \rangle\!\rangle$ denotes the subspace 
of ${\sf F}_2^n$ generated by the vector $1^n$ containing all 
ones.  The codes $\calC$ and $\calC^\perp$ correspond
to $\calC_2$ and $\calC_1$, respectively, in the 
Section \ref{sec-q-codes}; we have now added the requirement that 
$\calC_1^\perp = \calC_2$.
We follow MacWilliams et al.~\cite{McWSlTh}.  
They call a code {\it weakly self-dual} if
\begin{equation}
\langle\!\langle 1^n\rangle\!\rangle \subseteq \calC 
\subseteq \calC^\perp.
\end{equation}
Given a vector $v$ with even weight we need that the number of
$k$-dimensional
weakly self-dual codes for which $v \in \calC^\perp$ is 
independent of $v$.  In other words, the number of $k$-dimensional 
weakly self-dual codes $\calC$
contained in a given hyperplane $v^\perp$ is independent of $v$.

We apply Theorem 2.1 of Ref.~\cite{McWSlTh} (actually a stronger 
statement established in the proof).

Let $\sigma_{n,k,s}$ be the number of $k$-dimensional 
weakly self-dual codes $\calC_{[n,k]}$
that contain a given $s$-dimensional code $\calC_{[n,s]}$.  Then the
numbers $\sigma_{n,k,s}$ are independent of  the code 
$\calC_{[n,s]}$ that was chosen.

We separate the case 
$v \in \calC_{[n,k]} \subseteq \calC_{[n,k]}^\perp$ from
the case $v \in \calC_{[n,k]}^\perp \backslash \calC_{[n,k]}$.  
The number of $k$-dimensional weakly self-dual codes $\calC_{[n,k]}$ 
for which $v \in \calC_{[n,k]}$ is just $\sigma_{n,k,2}$, the number 
of codes containing
the 2-dimensional space $\langle\!\langle 1^n, v\rangle\!\rangle$.  
Next we consider pairs
$(\calC_{[n,k]},v)$ where $\calC_{[n,k]}$ is a $k$-dimensional weakly
self-dual code and 
$v \in \calC^\perp_{[n,k]} \backslash \calC_{[n,k]}$.
In this case $\calC_{[n,k]}$ and $v$ generate a $(k-1)$-dimensional 
weakly self-dual code $\calC_{[n,k+1]}$ containing the 2-dimensional 
space $\langle\!\langle 1^n,v\rangle\!\rangle$.  The number of 
choices for $\calC_{[n,k+1]}$ is 
$\sigma_{n,k+1,2}$.  Every code $\calC_{[n,k+1]}$ contains 
$2^k$ $k$-dimensional weakly self-dual codes of which $2^{k-1}$ do 
not contain the 2-dimensional space 
$\langle\!\langle 1^n, v\rangle\!\rangle$.  
Hence given a vector $v$
with even Hamming weight, the number of $k$-dimensional weakly 
self-dual codes contained in $v^\perp$ is independent of $v$.  
This is all that is needed to apply the greedy argument used 
to establish the Gilbert-Varshamov
bound.

The statement that there are codes meeting the Gilbert--Varshamov 
bound is that given a ratio $d/n$ (where $d$ denotes minimum 
distance), we may achieve a rate 
\begin{equation}
\textstyle (n-k)/n \geq 1-H_2\left(\frac{d}{n}\right).
\end{equation}
The redundancy $k/n$ satisfies ${k}/{n} \leq H_2({d}/{n})$,
so that the quantum codes achieve a rate
\begin{equation}
\textstyle R = (n-2k)/n \geq 1-2H_2\left(\frac{d}{n}\right).
\end{equation}
This function is plotted in Fig.~\ref{qcplots}.

\section{Quantum Channels}
\label{sec-q-channels}

In order to carry Shannon's theory of information to the quantum regime,
it is necessary to have some reasonable definition of a noisy quantum
channel.  We will define a quantum channel $W$ by a probability distribution 
$\calP$ on unitary transformations $U_W$ mapping 
$\calH_{\rm sig} \otimes \calH_{\rm  env}$.  For any pure input state
$\ket{x}$ the channel produces as output a mixed state by first obtaining 
an ensemble of states in $\calH_{\rm sig} \otimes \calH_{\rm env}$ 
by applying the transformation $U_W$ to $\ket{x}$ with 
probability distribution
$\calP$, and secondly tracing over $\calH_{\rm env}$.
While the initial state of $\calH_{\rm env}$ could be given by an
ensemble of states, it may also without loss of generality be taken
to be a fixed pure state, as the probability distribution given by
an ensemble of initial states may be absorbed into the probability
distribution on the unitary transformation $U_W$.  The probability
distribution could also be concentrated entirely in the inital mixed
state of $\calH_{env}$, and a fixed unitary transform $U$ be used, 
but this leads to a slightly less intuitive description
of the one quantum channel that we later discuss in detail.

Actual quantum channels are unlikely to produce output that differs
from the input exactly by the decoherence of at most $t$ qubits, and
thus are unlikely to be able to transmit quantum states perfectly using
this scheme.  However, if the average behavior of the channel results
in the decoherence of fewer than $t$ qubits, a channel may still be able 
to transmit quantum states very well.  A measure of the success of 
transmission of quantum states that has previously been successful
applied in quantum information theory is fidelity~\cite{bothSchu,BeBrScSmWo}.
In this paper, we define fidelity slightly differently
from the definition in Refs.~\cite{bothSchu}; we make this change as
these previous papers discuss
channels that transmit some distribution of
quantum states given {\it a priori}, whereas we want our channel to
faithfully transmit any pure input state.  Suppose that we have a
noisy channel $W$ that transmits quantum states in a Hilbert space
$\calH_{\rm sig}$.
We define the fidelity of the channel to be
\begin{equation}
\min_{ \ket{x} \in \calH_{\rm sig}} {\rm E} \bra{x} W \ket{x},
\end{equation}
where the expectation is taken over the output of the channel.
In other words, we are measuring the fidelity of transmission of the
pure state transmitted with least fidelity.  We could also measure
the fidelity of transmission of a typical state in $\calH_{\rm sig}$;
this average fideleity is a quantity which is closer to the previous 
definition, and may be more useful in some situations.

Assume that a channel $W$ transmits qubits with a fidelity of $F$ and is
that the decoherence process affects each qubit independently, i.e., each 
the decoherence of one qubit has no correlation with the decoherence of
any other qubit.  This would follow from the assumption that each qubit 
has a different environment, and this situation
corresponds to memoryless channels in classical information theory.
Then ${\rm E}_W \bra{x} W \ket{x} \geq F$ for every
state $\ket{x} \in \calH_2$.  If the output of our channel is a pure state,
our error-correction procedure $\calR_p\calR_f$ will be successful with 
probability equal to the length of the projection of the state onto 
the subspace of $\calH_2^n$ 
which results from decoherence of any $t$ or fewer qubits.  
Since the decoherence process for each qubit is independent, we can use
the binomial theorem to calculate the probability that the state $W^n\ket{y}$
is projected onto the correctable subspace of $\calH_2^n$,
where $\ket{y}$ is in our quantum code $\calC$.
We thus have a channel which transmits states $\ket{y}$ with fidelity
\begin{equation}
{\rm E} \bra{y} \calR_p \calR_f W^n \ket{y} \geq 
\sum_{j=0}^t {n \choose j} F^{n-j} (1-F)^j
\end{equation}
for all $\ket{y}$ in our quantum code $\calC$.  This quantity is close to 1
as long as $t/n > 1-F$.  Thus, if the fidelity $F$ for
each transmitted qubit is large enough, our quantum codes guarantee 
high fidelity transmission for our encoding of $k$ qubits.  
Our quantum codes will give good results for any channel $W$ that transmits 
states $\ket{y} \in \calH_2^n$ well enough that $W \ket{y}$ has an
expected projection of length at least
$1-\epsilon$ onto the subspace of $\calH_2^n$ obtained from $\ket{x}$
by the decoherence at most $t$ qubits.  Our encoding and decoding
schemes then give a channel on the Hilbert space $\calH_2^k$ which 
has fidelity $1-\epsilon$.  We will next use this observation to obtain 
an upper bound on the channel capacity of quantum channels.

An upper bound for the
amount of classical information carried by a quantum channel is
given by the Levitin--Holevo theorem \cite{levkh}.
If the output of the channel is a signal that has density 
matrix $\rho_a$ with probitility $p_a$, the Levitin--Holevo bound 
on the information content of this signal is
\begin{equation}
H(\rho) - \sum_a p_a H(\rho_a),
\end{equation}
where $\rho = \sum_a p_a \rho_a$ (the density matrix for the
ensemble of signals), and where $H(\rho) = -\Tr(\rho \log_2 \rho)$ is
the von Neumann entropy.  
Since quantum
information can be used to carry classical information, 
the Levitin--Holevo bound can be used
to obtain an upper bound for the rate of a
quantum error-correcting code.

Consider the following quantum channel discussed in 
Ref.~\cite{BeBrScSmWo}; this channel treats each qubit 
independently.  With probability $1-p$, a qubit is unchanged, 
corresponding to the identity transformation 
\renewcommand{\arraystretch}{.4}
$\left(\begin{array}{cc} \scriptstyle 1& \scriptstyle 0
\\ \scriptstyle 0 & \scriptstyle 1\end{array}\right)$.  
Otherwise, with each possibility having probability
$p/3$, the qubit is acted on by the unitary transformation 
corresponding to one of the three matrices:
$\left(\begin{array}{cc}  \scriptstyle 0& \scriptstyle 1\\  
\scriptstyle 1& \scriptstyle 0\end{array}\right)$,
$\left(\begin{array}{cc}  \scriptstyle 1& 
\scriptstyle \phantom{-}0\\  
\scriptstyle 0& \scriptstyle -1\end{array}\right)$, or
$\left(\begin{array}{cc}  \scriptstyle 
\phantom{-}0& \scriptstyle 1\\  
\scriptstyle -1& \scriptstyle 0\end{array}\right)$.
\renewcommand{\arraystretch}{1}
That is, each of the following possibilities
has probability $p/3$: the qubit is negated, 
or its phase is changed, or it is both negated and its phase 
is changed.  If $t/n > p+\epsilon$ for $\epsilon>0$, 
the length projection of the output of this channel
onto the subspace of $\calH_2^n$ with at most $t$ errors
approaches 1 as $n$ grows, so
the quantum error-correcting codes
given earlier in this paper guarantee high
fidelity.  This channel can alternatively
be described as transmitting a qubit error-free with
probability $1-\frac{4}{3}p$, and producing a random quantum 
state with probability $\frac{4}{3}p$.  This description
shows that the entropy of the
output of the channel is at least $H_2(\frac{2}{3}p)$, so by the 
Levitin--Holevo theorem an upper bound on the classical information
capacity of this channel is $1-H_2(\frac{2}{3}p)$. 
This bound is plotted in Fig.~\ref{qcplots}.  For this channel,
the bound is achievable for classical information, but we believe
it is unlikely to be tight for quantum information.

Another question that has been studied is: how much entanglement
can be transmitted over a quantum channel \cite{BeBrScSmWo}.  Since
any means of transmitting quantum states with high fidelity
can also be used to
transmit entanglement, upper bounds for entanglement transmission
also apply to the quantum information capacity of a quantum channel.  
For the above channel, the upper bound proved in 
Ref.~\cite{BeBrScSmWo} is $H_2(\frac{1}{2}+\sqrt{p(1-p)})$
for $p < \frac{1}{2}$ and 0 if $p \geq \frac{1}{2}$.  This bound is 
also plotted in Fig.~\ref{qcplots}.

\acknowledgments

We would like to thank Peter Winkler for helpful discussions on quantum
error-correcting codes, and David DiVincenzo for advice on the
presentation of these results.

\begin{figure}
\caption{The solid line shows the asymptotic rate $R$ of our quantum 
codes versus the error rate of the channel $t/n$.  Two upper bounds 
for this quantity are also plotted: the Levitin--Holevo upper bound 
with a dashed line and the entanglement upper bound with a dotted 
line.}
\label{qcplots}
\end{figure}

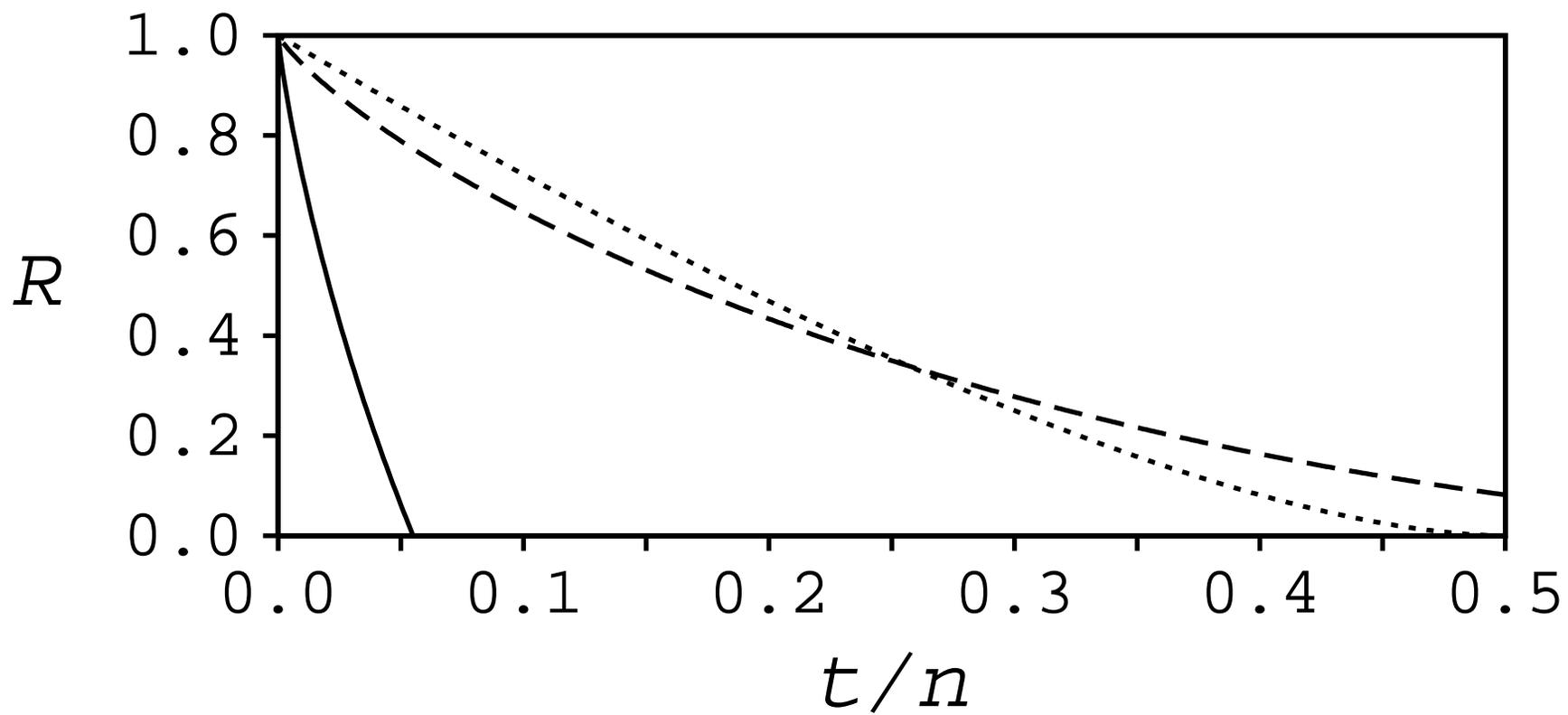
\end{document}